\documentclass{article}
\usepackage{spconf,amsmath,amssymb,graphicx,hyperref}
\usepackage{booktabs}
\usepackage[table]{xcolor}
\usepackage{tabularx}
\usepackage{array}

\title{WEBEXPERT: DOMAIN-AWARE WEB AGENTS WITH CRITIC-GUIDED EXPERT EXPERIENCE FOR HIGH-PRECISION SEARCH}

\name{Yuelin Hu$^{1}$, Zhengxue Cheng$^{1}$, Ronghua Wu$^{2}$, Qunshan Gu$^{2}$, Hongwei Hu$^{2}$, Wei Liu$^{3}$, Qiao Liang$^{4}$, Li Song$^{1}$}
\address{$^{1}$ Shanghai Jiao Tong University\\
$^{2}$ Ant Group\\
$^{3}$ Shanghai Maritime University\\
$^{4}$ Shanghai Tongji University\\
\texttt{\{huyuelin51717221,zxcheng,song\_li\}@sjtu.edu.cn}\\
\texttt{\{r.wu,guqunshan.gqs,Hongwei.huhw\}@antgroup.com}\\}

\begin{document}
%
\maketitle

\begin{figure*}[t]
\centering
\includegraphics[width=0.9\textwidth]{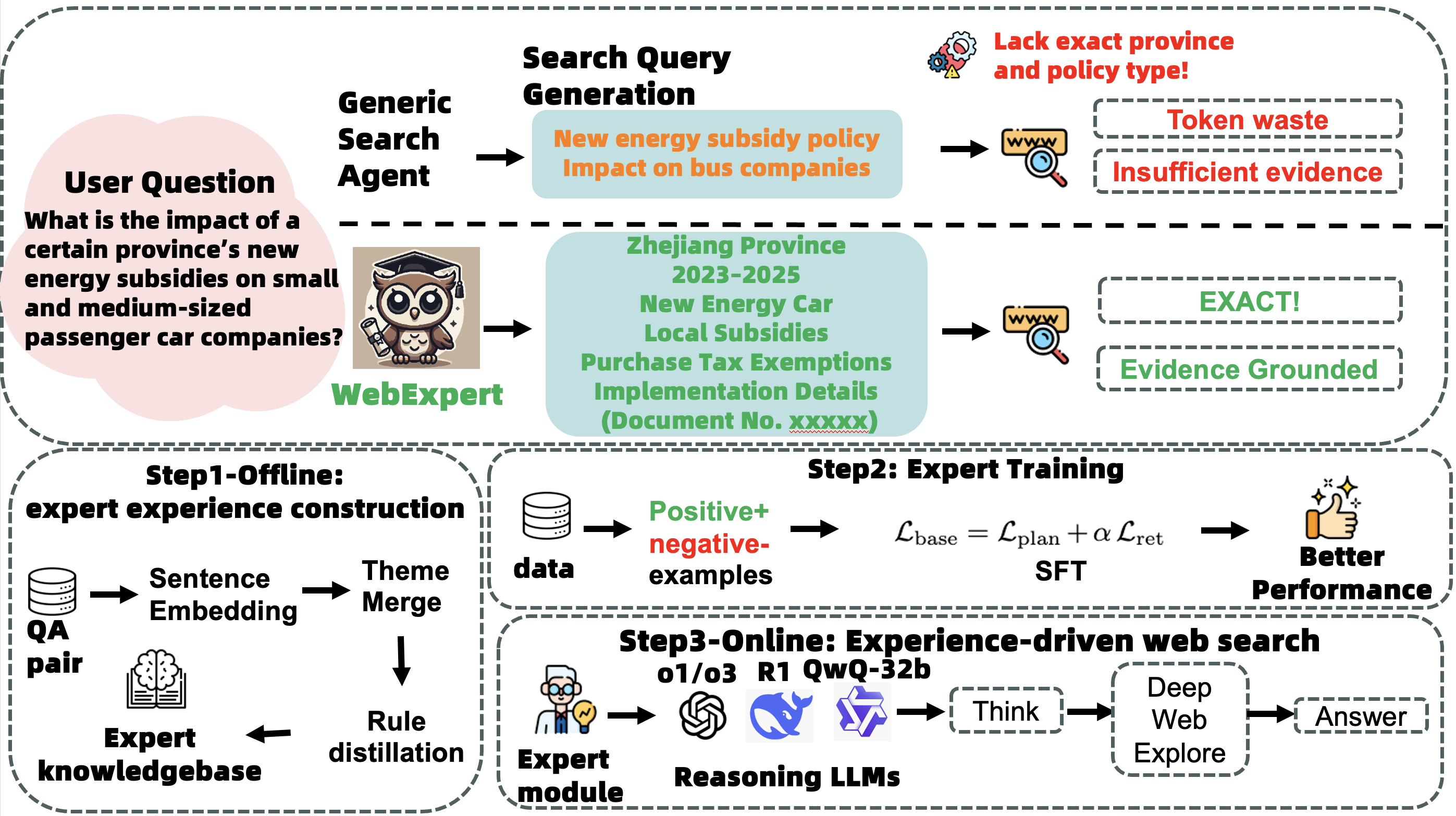}
\caption{Overview of WebExpert. Top: compared to a generic search agent, WebExpert grounds queries with exact facets (e.g., province, time span, policy type), yielding evidence-grounded retrieval rather than token waste and insufficient evidence. Bottom: the three-step pipeline. Step 1 (offline): construct an expert experience base from QA pairs and curated sources via sentence embedding, topic merging, and rule distillation. Step 2 (training): experience-aware supervised fine-tuning (SFT) optimizes planning and retrieval. Step 3 (online): an experience-driven module conditions reasoning large language models (LLMs) to think, perform deep web exploration, and generate answers.}
\label{fig:overview}
\end{figure*}

\begin{abstract}
Specialized web tasks in finance, biomedicine, and pharmaceuticals remain challenging due to missing domain priors: queries drift, evidence is noisy, and reasoning is brittle. We present WebExpert, a domain-aware web agent that we implement end-to-end, featuring: (i) sentence-level experience retrieval with topic merging and rule distillation, (ii) schema-light facet induction that bootstraps time/region/policy/industry facets from weak supervision instead of static hand-written lexicons, and (iii) preference-optimized planning that jointly improves query planning and retrieval via pairwise preference learning alongside a coverage-aware objective. At inference, a lightweight experience gate biases decoding toward active facets with fallback under low-retrieval confidence. On GAIA, GPQA, HLE, and WebWalkerQA, WebExpert improves Answer Exact Match (EM) by 1.5--3.6 pp over the strongest browsing baseline and reduces page hops. Analysis shows consistent gains and ablations on retrieval, topic merging, facet induction, and preference-aware training. Our code is available at \url{https://github.com/huyuelin/WebExpert}. 
\end{abstract}
\section{Introduction}
\label{sec:intro}

Web browsing agents have shown strong results on open-ended tasks, yet their effectiveness drops in domain-specific scenarios (e.g., credit approval in finance, clinical guidance in biomedicine). Without expert priors, agents formulate off-target queries, wander to irrelevant pages, and miss evidence. In practice, domain practitioners attend to contextual factors such as seasonality, regional regulations, and domain-specific granularity; generic agents rarely do.

Figure~\ref{fig:overview} provides an overview contrasting a generic search agent with our WebExpert and summarizes the three-step pipeline that underpins our system.

We present \textbf{WebExpert}, a domain-aware web agent that integrates an \emph{expert experience module} before deep browsing. The module retrieves domain experiences and generates domain-grounded queries that steer our in-house browsing controller. Our key idea is a \emph{critic-guided extraction chain} that converts annotated data and expert materials into reusable sentence-level experiences, merged into concise rules that generalize within a domain.

\textbf{Contributions} (concise). (i) We formulate domain-aware web browsing via a critic-guided extraction chain that injects sentence-level expert priors to steer query semantics along domain-relevant facets. (ii) We present a practical pipeline from sentence extraction and dense embedding to topic clustering/merging and rule distillation (Uniform Manifold Approximation and Projection, UMAP; Hierarchical Density-Based Spatial Clustering of Applications with Noise, HDBSCAN; and BERTopic) \cite{umap,hdbscan,bertopic}. (iii) We introduce schema-light facet induction that automatically induces facet vocabularies from weak supervision and corpus statistics, reducing manual schema dependence. (iv) We propose experience-conditioned planning with coverage-aware supervised fine-tuning (SFT), retrieval margin, and preference optimization, improving precision beyond generic Retrieval-Augmented Generation (RAG) \cite{rag,retrollm,selfrag,unigen}. (v) On GAIA/GPQA/HLE and WebWalkerQA, WebExpert yields consistent 1.5--3.6 pp EM gains over the strongest browsing baseline, with fewer page hops.

\section{Related Work}
\label{sec:related}

\textbf{Web agents and deep research.} Large reasoning models (LRMs) integrated with search and browsing have shown strong capabilities in complex tasks. Reason-then-search systems (e.g., search-o1 \cite{searcho1}) couple agentic retrieval with in-document reasoning to iteratively refine external knowledge. Recent pipelines (e.g., \cite{webthinker}) synthesize broad web search with deeper on-page exploration, enabling navigation across multi-step webpages while maintaining coherent reasoning. These frameworks achieve notable gains in GPQA/GAIA/WebWalkerQA/HLE, yet rely on generic priors and may drift in specialized domains.

\textbf{Retrieval-Augmented Generation.} Retrieval-Augmented Generation (RAG) \cite{rag} and its variants \cite{retrollm,selfrag,unigen} improve language models via retrieval augmentation and self-critique. However, RAG quality depends critically on query semantics, ranking, and denoising; domain-specialized priors (policy, region, L2 industry) are rarely injected in a structured way. Our approach complements RAG by distilling sentence-level, facetized experiences that directly bias query planning while remaining \emph{schema-light}. 

\vspace{-0.355cm}
\section{Method}
\label{sec:method}

\subsection{Problem Setup}
We consider domain-specific web tasks where an agent must generate search queries, browse the web, and synthesize an answer \(a\) for a question \(q\). We assume access to a curated \emph{expert experience base} \(\mathcal{E}=\{r_i\}_{i=1}^{N}\) of sentence-level rules distilled from expert corpora. Let \(\mathcal{E}^{(k)}\) denote the top-\(k\) retrieved experiences for \(q\) (see Sec.~\ref{sec:inference}). The overall mapping follows a reasoning-with-experiences paradigm:
{\footnotesize
\begin{equation}
\begin{aligned}
P(\mathcal{R}, a \mid q, \mathcal{E}^{(k)})
&= \prod_{t=1}^{T_r} P\big(\mathcal{R}_t \mid \mathcal{R}_{<t}, q, \mathcal{E}^{(k)}_{\le t}, \mathcal{D}_{<t}\big) \\
&\quad \cdot \prod_{t=1}^{T_a} P\big(a_t \mid a_{<t}, \mathcal{R}, q\big),
\end{aligned}
\label{eq:overall}
\end{equation}.}
where \(\mathcal{R}\) is the reasoning chain, \(\mathcal{D}_{<t}\) are retrieved web documents prior to step \(t\), and \(\mathcal{E}_{\leq t}\) denotes the subset of retrieved experiences used up to step \(t\).

\subsection{Critic-Guided Expert Experience Extraction}

\textbf{(1) Question harvesting and canonicalization.} We collect tuples \(\mathcal{D}=\{(q_i, a_i, \mathcal{R}_i, \mathcal{C}_i)\}\) (question, final answer, optional reasoning chain, and citations). Surface forms of questions are normalized via paraphrase mining and schema-free delexicalization to get canonical intents \(\tilde{q}_i\).

\textbf{(2) QA-level multi-view clustering.} We compute representations for both questions and answers, e.g., \(\mathbf{u}_i=f_Q(q_i)\), \(\mathbf{v}_i=f_A(a_i)\), and optionally a co-encoded pair \(\mathbf{w}_i=f_{QA}(q_i,a_i)\). Multi-view density clustering (HDBSCAN/BERTopic or spectral) groups QA tuples under a similarity
\[ s\big((q,a),(q',a')\big) = \lambda_1\,\langle \mathbf{u},\mathbf{u}'\rangle + \lambda_2\,\langle \mathbf{v},\mathbf{v}'\rangle + \lambda_3\,\langle \mathbf{w},\mathbf{w}'\rangle, \]
with soft assignments to allow overlapping intents. This QA-first view captures semantically similar problems even when answers differ in granularity.

\textbf{(3) Evidence aggregation and de-duplication.} For each cluster \(T_m\), we aggregate answers and mined rationales, retain top-ranked pages/quotes using the BM25 ranking function (BM25) and dense retrieval, and apply Maximal Marginal Relevance (MMR) \cite{mmr} for diversity. Source-level diversity and quote-level de-duplication reduce redundancy and noise.

\textbf{(4) Contradiction-aware summarization (DeepSeek-R1\footnote{DeepSeek-R1: a publicly available large reasoning model used for contradiction-aware summarization and synthesis.}).} We prompt a contradiction- and uncertainty-aware summarizer (e.g., DeepSeek-R1) with the set of answers/rationales and citations in \(T_m\) to produce a concise rule \(r_m\) that includes: conditions (assumptions), core guidance, edge cases, and known failure modes. A lightweight entailment/consistency check filters self-contradictory statements; majority-consistent claims are preferred while minority views are either folded into caveats or flagged.

\textbf{(5) Facetization and normalization.} Each rule is facetized into \(g_m=(\textit{time},\textit{region},\textit{policy},\textit{L2 industry})\) by first filtering high-frequency domain terms (e.g., ``CFA Institute'' for finance, ``FDA'' for biomedicine) via corpus statistics as facet candidates, then refining with shallow taggers and LLM disambiguation. We normalize time ranges, geographic names, and policy references, and attach metadata (coverage, confidence, provenance).

\textbf{(6) Continuous refresh and versioning.} The experience base is maintained as a versioned store with warm-start clustering and local merges, preserving stable rule identifiers and citation sets while enabling streaming updates as new QA/rationale evidence arrives.

Formally, let clustering yield \(\mathcal{T}=\{T_m\}_{m=1}^{K}\) with \(T_m\subset\mathcal{D}\). For each \(T_m\), the summarizer \(h(\cdot)\) produces a rule and citations
\[ (r_m, c_m, g_m) = h\big(T_m\big), \quad c_m \subseteq \bigcup_{(q,a,\mathcal{R},\mathcal{C})\in T_m} \!\mathcal{C}. \]
The resulting experience base is \(\mathcal{E}=\{(r_m,c_m,g_m)\}_{m=1}^{K}\). When answers are unavailable, we fallback to sentence-level extraction followed by the same consolidation, preserving compatibility with prior pipelines \cite{umap,hdbscan,bertopic}. Table~\ref{tab:exp_cluster_example} is a concrete example of how clustered QAs are distilled into expert experiences.

\begin{table*}[t]
\centering
\footnotesize
\setlength{\tabcolsep}{3pt}
\renewcommand{\arraystretch}{1.1}
\caption{From clustered QAs to distilled expert experience: asset correlation and diversification (GPQA - Finance).}
\label{tab:exp_cluster_example}
\begin{tabularx}{\textwidth}{@{}X X X X@{}}
\toprule
\rowcolor{gray!10}
\textbf{QA Example 1} & \textbf{QA Example 2} & \textbf{QA Example 3} & \textbf{Distilled Expert Experience} \\
\midrule
\textbf{Q:} When is diversification most effective in portfolio risk management?

\textbf{A:} Diversification is most effective when portfolio assets are uncorrelated.

\textit{Source:} Investopedia, CFAI
& 
\textbf{Q:} Does asset correlation affect diversification benefits in investing?

\textbf{A:} Yes, higher correlation among assets reduces the risk reduction benefit of diversification.

\textit{Source:} BlackRock, Morningstar
& 
\textbf{Q:} How do correlations impact portfolio volatility?

\textbf{A:} Lower asset correlations lead to lower overall portfolio volatility due to better risk spreading.

\textit{Source:} Corp Finance, CFAI
& 
\textbf{Rule ($r_m$):} Diversification is most impactful when portfolio assets exhibit low or negative correlation; in such scenarios, overall risk and volatility are minimized.

$\bullet$ \textit{Time:} Ongoing principle 

$\bullet$ \textit{Region:} Universal context \\


\bottomrule
\end{tabularx}
\end{table*}

\subsection{Inference}
\label{sec:inference}
During inference, WebExpert prepends an expert experience module to our deep browsing controller:
\begin{enumerate}
\item \textbf{Experience retrieval}: compute 

$\mathcal{E}^{(k)}=\mathrm{Top}\text{-}k\{ s(f(q), f(r)) : r\in\mathcal{E}\}$, where $s(\mathbf{u},\mathbf{v})=\langle\mathbf{u},\mathbf{v}\rangle/(\lVert\mathbf{u}\rVert\lVert\mathbf{v}\rVert)$.
\item \textbf{Domain-grounded query generation}: produce a multi-query plan \(\mathbf{z}=(z_1,\ldots,z_M)\) conditioned on \(q\) and \(\mathcal{E}^{(k)}\), with an experience gate that biases decoding toward active facets. The gate's retrieval confidence is computed as the average cosine similarity of top-\(k\) experiences (threshold \(\theta=0.3\), calibrated on validation set); when confidence \(<\theta\), it falls back to generic query generation to avoid over-constraint:
{\footnotesize
\begin{equation}
\begin{aligned}
P(\mathbf{z}\mid q, \mathcal{E}^{(k)})
&= \prod_{j=1}^{M} P\big(z_j\mid z_{<j}, q, \mathcal{E}^{(k)}\big).
\end{aligned}
\label{eq:querygen}
\end{equation}.}
\item \textbf{Deep browsing}: feed \(\mathbf{z}\) to our search-and-browse controller, interleaving retrieval \(\mathcal{D}\) and reasoning \(\mathcal{R}\) to produce \(a\) per Eq.~\eqref{eq:overall}.
\end{enumerate}

\subsection{SFT and Training Objectives}
We fine-tune QwQ-32B to optimize an \textit{experience-aware} objective that jointly encourages (a) generating domain-grounded queries consistent with retrieved facets and (b) preferring experience-relevant rules during retrieval. The planner is trained with a token objective weighted by facet alignment:
{\footnotesize
\begin{equation}
\mathcal{L}_{\mathrm{plan}}= - \sum_{q}\sum_{t} w(y_t;\,\phi(\mathcal{E}^{(k)}))\, \log \pi_{\theta}\big(y_t\mid y_{<t}, q, \mathcal{E}^{(k)}\big),
\label{eq:sft}
\end{equation}.}
where $\phi(\mathcal{E}^{(k)})$ maps the retrieved experiences to facet indicators (time, region, policy, industry) and associated keywords, and $w(\cdot)$ up-weights tokens that activate these indicators while down-weighting off-facet tokens. Beyond Eq.~\ref{eq:sft}, we add retrieval-margin, coverage, and preference terms to encourage selecting high-quality experiences and facet coverage. In particular, we optimize a contrastive retrieval objective:
{\footnotesize
\begin{equation}
\mathcal{L}_{\mathrm{ret}} = - \sum_{q} \log \frac{\exp\big( s\big(f(q), f(r^+)\big)/\tau \big)}{\exp\big( s\big(f(q), f(r^+)\big)/\tau \big) + \sum\limits_{r^-} \exp\big( s\big(f(q), f(r^-)\big)/\tau \big)} ,
\label{eq:ret}
\end{equation}.}
where $r^+$ is a positive experience aligned with $q$, $r^-$ are hard negatives, $f(\cdot)$ are the encoders used in retrieval, $s(\cdot,\cdot)$ is cosine similarity, and $\tau$ is a temperature.

\section{Experiments}
\label{sec:exp}

\subsection{Setup}
\textbf{Datasets.} We evaluate on GAIA, GPQA, HLE. Each includes open-domain and domain-focused subsets; we report overall results. We additionally evaluate on WebWalkerQA, a benchmark for multi-step web browsing and grounded question answering. WebWalkerQA includes hundreds of tasks across real-world domains and requires page navigation with evidence citation. 


\textbf{Metrics.} (i) \textit{Exact Match (EM)} and \textit{F1 score (F1)}; (ii) \textit{Query Precision@3} (QP@3): proportion of generated queries that retrieve on-topic evidence; (iii) \textit{Page Hops} (hops per solved example); (iv) \textit{Evidence normalized Discounted Cumulative Gain at 10 (nDCG@10)} over cited pages; (v) \textit{Leakage stress tests}: entity-randomized EM, time-shifted EM, and template-remix EM.


\subsection{Training Details}
We train with $\sim$12k preference-aligned pairs curated from expert rules and browsing trajectories: positives emphasize facet-aligned plans; negatives suppress off-facet or redundant plans. Full-parameter fine-tuning of QwQ-32B uses Pai-Megatron-Patch \cite{pai_patch} (lr=\(1e^{-5}\), cosine decay, \(\beta_2=0.98\)). Validation uses QP@3 and EM on held-out GAIA items; early stopping selects the best checkpoint. For \(\mathcal{L}_{\mathrm{ret}}\), negatives are sampled via hard-negative mining from top-64 Facebook AI Similarity Search (FAISS) candidates excluding positives (score margin within 0.05), refreshed every epoch.

\subsection{Evaluation Protocol and Statistical Rigor}
QP@3 counts a query correct if at least one of its top-3 retrieved pages contains answer-bearing evidence (LLM-as-judge + strict match). Page Hops counts unique page visits until final answer. nDCG@10 is computed over cited URLs ranked by the agent. All systems use Bing (US-EN), top-10 retrieval, temp 0.7, max tokens 32k.

\subsection{Main Results}
Table~\ref{tab:main} shows that WebExpert outperforms strong baselines on Answer EM across datasets, with consistent gains under standardized settings. We additionally include WebWalkerQA under the same protocol for completeness.

\begin{table}[t]
\centering
\scriptsize
\setlength{\tabcolsep}{1pt}
\resizebox{\columnwidth}{!}{%
\begin{tabular}{lcccc}
  \toprule
  Method & GAIA EM & GPQA EM & HLE EM & WebWalkerQA EM \\
  \midrule
  QwQ-32B (Direct) & 13.6$\,\pm\,$0.7 & 43.4$\,\pm\,$0.6 & 5.4$\,\pm\,$0.8 & 3.1$\,\pm\,$1.9 \\
  RAG-QwQ-32B & 32.0$\,\pm\,$0.6 & 64.6$\,\pm\,$0.7 & 7.2$\,\pm\,$0.7 & 31.2$\,\pm\,$1.7 \\
  Search-o1-32B & 39.8$\,\pm\,$0.6 & 67.2$\,\pm\,$0.6 & 10.8$\,\pm\,$0.6 & 34.1$\,\pm\,$1.6 \\
  WebThinker-32B-Base & 44.7$\,\pm\,$0.5 & 68.7$\,\pm\,$0.6 & 13.0$\,\pm\,$0.6 & 41.9$\,\pm\,$1.5 \\
  \midrule
  WebExpert (ours) & \textbf{46.2}$\,\pm\,$0.6$^{\dagger}$ & \textbf{70.2}$\,\pm\,$0.5$^{\dagger}$ & \textbf{14.5}$\,\pm\,$0.6$^{\dagger}$ & \textbf{43.7}$\,\pm\,$1.2$^{\dagger}$ \\
  WebExpert+SFT & \textbf{47.7}$\,\pm\,$0.5$^{\dagger}$ & \textbf{71.9}$\,\pm\,$0.5$^{\dagger}$ & \textbf{16.6}$\,\pm\,$0.5$^{\dagger}$ & \textbf{46.3}$\,\pm\,$1.1$^{\dagger}$ \\
  \bottomrule
  \end{tabular}}
  \caption{Main results (mean$\pm$std over 3 seeds $\times$ 2 traces). EM in \%. $\dagger$: $p{<}0.01$ vs. strongest browsing baseline by paired t-test.}
  \label{tab:main}
  \end{table}

\subsection{Query Quality}
We observe QP@3 increases from 49.3 (WebThinker) to 58.2 (WebExpert) and 61.8 (WebExpert+SFT). Page hops drop from 8.1 to 5.6 and 5.2, respectively. Evidence nDCG@10 improves by 4--6 points across datasets.

\subsection{Ablation Studies}
We ablate components on GAIA. The ablation uses a stratified subset with different judging thresholds to accelerate iterations; therefore absolute EM is not directly comparable to Table~\ref{tab:main} and we focus on relative trends.
\begin{table}[t]
  \centering
\small
\setlength{\tabcolsep}{6pt}
\begin{tabular}{lcc}
\toprule
Variant & EM (\%) & QP@3 (\%) \\
\midrule
WebExpert (full) & 47.7 & 61.8 \\
\quad w/o SFT & 46.2 & 58.2 \\
\quad w/o topic merging & 44.1 & 59.1 \\
\quad w/o sentence-level embedding & 45.7 & 56.0 \\
\quad top-\(k\)=1 (vs. 5) & 41.2 & 57.1 \\
\bottomrule
\end{tabular}
\caption{Ablation on GAIA. Sentence-level embeddings and SFT contribute most; retrieving top-5 experiences balances precision and coverage.}
\label{tab:ablation}
\end{table}

\section{Acknowledgment}
This work was partly supported by the NSFC (62431015, 62571317, 62501387), the Fundamental Research Funds for the Central Universities, Shanghai Key Laboratory of Digital Media Processing and Transmission under Grant 22DZ2229005, 111 project BP0719010.


\section{Conclusion}
\label{sec:conclusion}
We proposed WebExpert, a critic-guided, domain-aware web agent that retrieves expert experiences to ground query generation before deep browsing. Experiments on GAIA, GPQA, and HLE show consistent 1.5--3.6 pp gains and improved efficiency. Our analysis highlights the importance of sentence-level retrieval, topic merging, and SFT for domain fidelity.

\vfill\pagebreak

\end{document}